\documentclass[aps,pre,twocolumn,groupedaddress,superscriptaddress,showpacs,nofootinbib,notitlepage]{revtex4-1}

\usepackage{makecell} 
\usepackage{xurl}
\usepackage[hidelinks]{hyperref}
\usepackage{booktabs}       
\usepackage{amsfonts}       
\usepackage{amssymb}
\usepackage{mathtools}
\usepackage{nicefrac}       
\usepackage{microtype}      
\usepackage{amsmath}
\usepackage{xcolor,colortbl}
\usepackage{hhline}
\usepackage{bbm}
\usepackage[draft]{changes}
\usepackage{tabularx}
\usepackage{amsthm}
\usepackage{changes}
\usepackage{comment}
\usepackage{color,soul}
\usepackage{xspace}
\usepackage{multirow}
\usepackage{etoolbox}
\usepackage{tcolorbox}
\patchcmd{\section}
  {\centering}
  {\raggedright}
  {}
  {}
\patchcmd{\subsection}
  {\centering}
  {\raggedright}
  {}
  {}

\def\ie{\textit{i.e.}}

\makeatletter
\newcommand{\lowersim}[2]{%
  \sbox\z@{$#1<$}%
  \raisebox{-\dimexpr\height-\ht\z@}{$\m@th#1#2$}%
}
\makeatother

\makeatletter
\DeclareRobustCommand\onedot{\futurelet\@let@token\@onedot}
\def\@onedot{\ifx\@let@token.\else.\null\fi\xspace}

\begin{document}
\title{Control of Medical Digital Twins with Artificial Neural Networks}

\author{Lucas B\"{o}ttcher}
\thanks{\href{mailto:l.boettcher@fs.de}{l.boettcher@fs.de}}
\affiliation{Dept.~of Computational Science and Philosophy, Frankfurt School of Finance and Management, Frankfurt am Main, 60322, Germany}
\affiliation{Laboratory for Systems Medicine, Department of Medicine, University of Florida, Gainesville, FL, USA}
\author{Luis L. Fonseca}
\affiliation{Laboratory for Systems Medicine, Department of Medicine, University of Florida, Gainesville, FL, USA}
\author{Reinhard C. Laubenbacher}
\affiliation{Laboratory for Systems Medicine, Department of Medicine, University of Florida, Gainesville, FL, USA}
\date{\today}
\begin{abstract}
The objective of personalized medicine is to tailor interventions to an individual patient's unique characteristics. A key technology for this purpose involves medical digital twins, computational models of human biology that can be personalized and dynamically updated to incorporate patient-specific data collected over time. Certain aspects of human biology, such as the immune system, are not easily captured with physics-based models, such as differential equations. Instead, they are often multi-scale, stochastic, and hybrid. This poses a challenge to existing model-based control and optimization approaches that cannot be readily applied to such models. Recent advances in automatic differentiation and neural-network control methods hold promise in addressing complex control problems. However, the application of these approaches to biomedical systems is still in its early stages. This work introduces dynamics-informed neural-network controllers as an alternative approach to control of medical digital twins. As a first use case for this method, the focus is on agent-based models, a versatile and increasingly common modeling platform in biomedicine. The effectiveness of the proposed neural-network control method is illustrated and benchmarked against other methods with two widely-used agent-based model types. The relevance of the method introduced here extends beyond medical digital twins to other complex dynamical systems.
\end{abstract}
\maketitle
%
%
The ultimate goal of personalized medicine is to identify interventions that can preserve or restore an individual's health by taking into account their unique personal characteristics. Computational models known as medical digital twins play an important role in realizing this goal~\cite{bjornsson2020digital,laubenbacher2021using,masison2021modular}. Medical digital twins are designed to incorporate the most recent personal health data, offering guidance for the application of optimal interventions. 

Developing medical digital twins is challenging as the underlying models must capture biological mechanisms operating at various spatial and temporal scales, such as the impact of drugs on both the intracellular scale and the larger organ or organism scale. Depending on the specific application, digital twins may also need to account for stochastic effects. Consequently, developing high-fidelity medical digital twins often necessitates the incorporation of high-dimensional, multiscale, stochastic computational models. Given the difficulties associated with representing these intricacies using equation-based approaches, alternative model types, such as agent-based models (ABMs), frequently serve as the foundation for medical digital twins~\cite{oremland2016computational,masison2021modular,an2021agent,ribeiro2022multi,joslyn2022virtual,joslyn2022concomitant,ribeiro2023covid,budak2023optimizing,west2023agent}. In biomedical applications, agents in an ABM represent biological entities such as cells in a tissue or microbes in a biofilm~\cite{weston2015agent,bauer2017bacarena,lin2018gutlogo,archambault2021understanding}. The behavior of agents is usually described by stochastic rules, which allow them to navigate heterogeneous spatial environments and interact with each other. Since ABMs are intuitive and easily implementable computational models, they are accessible to domain experts without extensive computational modeling knowledge. They find applications in various medical scenarios, including studies on the immune system, tumor growth, and treatment development~\cite{oremland2016computational,masison2021modular,an2021agent,ribeiro2022multi,joslyn2022virtual,joslyn2022concomitant,ribeiro2023covid,budak2023optimizing,west2023agent}. A major drawback of using ABMs and other non-equation-based models is that the technology underpinning their analysis and use is largely missing, including identifiability of parameters, practical sensitivity analysis methods, forecasting algorithms, and control and optimization tools. The results presented in this paper can be viewed as a contribution to the development of mathematical tools appropriate for the model types likely to be used for medical digital twins, as the use of this technology continues to expand. 

Designing effective treatments using ABM-based models is computationally challenging, due to the typically large state space of ABMs and the associated ``curse of dimensionality''. Additionally, while optimal control theory methods are well-established for ordinary differential equation (ODE) models in engineering~\cite{ogata2009modern,aastrom2021feedback}, they are not readily applicable to complex hybrid models. Hence, identifying optimal controls using ABMs often relies on ad hoc methods. This issue extends beyond medicine to digital twin systems in various domains. As highlighted in the 2023 report on fundamental research gaps for digital twins by the National Academies of Engineering, Science, and Medicine, there are currently no general solutions available to address this challenge~\cite{national2023foundational}.

While progress has been made in connecting approaches from control theory with biomedical ABMs~\cite{an2017optimization}, these methods are often only applicable to ODE metamodels~\cite{nardini2021learning,fonsecametamodels2024} and not to the original ABMs. Already in 1971, Alexey Ivakhnenko commented in his work on polynomial neural networks on the challenges associated with the application of control theory to complex systems~\cite{ivakhnenko1971polynomial}: ``Modern control theory, based on differential equations, is not an adequate tool for solving the problems of complex control systems. It is necessary to construct differential equations to trace the input-output paths, that is, to apply a deductive deterministic approach. But it is impossible to use this approach for complex systems because of the difficulty in finding these paths.'' His work is nowadays regarded as a foundational contribution to the field of deep learning~\cite{schmidhuber2015deep}. 

In parallel to the development of ABMs that provide the core components of many medical digital twins, deep artificial neural networks (ANNs) have become a common approach for deploying general function approximators in various machine-learning tasks, including dynamical system identification~\cite{wang1998runge,DBLP:conf/nips/ChenRBD18,lagergren2020biologically,fronk2023interpretable} and control~\cite{bottcher2022ai,asikis2022neural,bottcher2022near,mowlavi2023optimal,bottcher2023control,bottcher2023gradient}. Building upon these recent advances in neural-network control and automatic differentiation, in this paper, we develop ANN control approaches to effectively steer ABMs toward desired target states. We examine the performance of the proposed controllers with respect to recently developed metamodeling methods~\cite{fonsecametamodels2024}.

To illustrate the effectiveness of the proposed control approaches, we apply them to two paradigmatic ABMs: (i) a resource-dependent predator-prey system~\cite{wilensky1997netlogo,wilensky1999netlogo}, and (ii) a regulated metabolic network. The predator-prey system that we consider was originally used to model the interaction between grass, sheep, and wolves, but these types of models also find widespread applications in biomedicine~\cite{may1975nonlinear,faust2012microbial,cortez2014coevolution,joseph2020compositional}. In this context, certain immunological processes resemble those observed in ecological systems, where pathogens act as predators, preying on host cells or consuming host resources. Conversely, host cells can also function as predators and pathogens as prey. Unlike typical predators that consume prey to sustain their populations, immune cells do not rely on pathogen phagocytosis for growth. However, contact with pathogens tends to attract more immune cells, a phenomenon encoded similarly using mass-action kinetics. For example, when {SARS-CoV-2} enters the human airways, it infects epithelial cells, while neutrophils and macrophages target the pathogen. Similarly, in cases of fungal infections like Aspergillosis~\cite{oremland2016computational,ribeiro2022multi}, the pathogen scavenges iron from the host to support its growth. Meanwhile, host immune cells target the fungus and lock iron intracellularly, prompting the fungus to invade nearby blood vessels to access iron in hemoglobin. This scenario mirrors a predator-prey dynamic, with iron serving as the resource exploited by the fungus, acting as the prey. 

The metabolic pathway model that we consider in this paper describes the synthesis of two end products originating from a shared precursor. One of these end products serves a dual role: it inhibits the initial reaction in the pathway while stimulating metabolic flux towards the second end product. Such metabolic processes are prevalent in branched pathways, such as amino acid biosynthesis.

Overall, our work contributes to expanding the applicability of ABMs across fields and offers a new perspective on controlling complex biomedical systems.
\section*{Results}
\subsection*{Predator-prey model}
\begin{figure}
    \centering
    \includegraphics[width=0.49\textwidth]{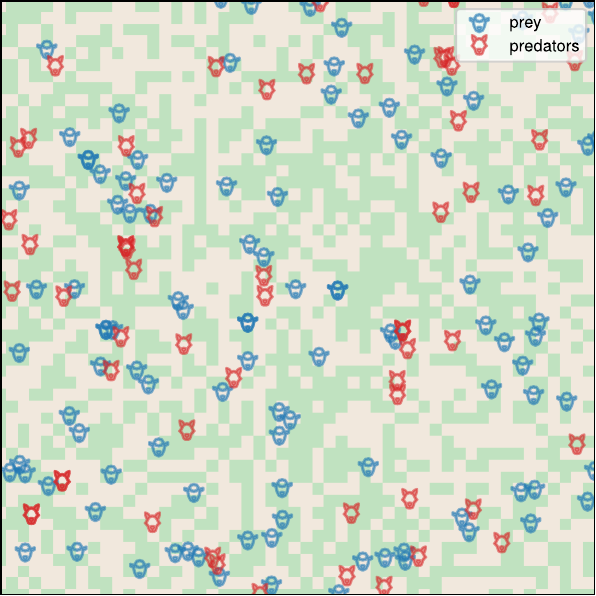}
    \caption{Predator-prey ABM dynamics. A snapshot of a three-species predator-prey ABM simulation with $51\times51$ grid cells. Green and light brown grid cells represent nutrient-rich and nutrient-poor regions, respectively.}
    \label{fig:abm}
\end{figure}
We first consider an extended predator-prey system with three species $A$, $B$, and $C$~\cite{may1975nonlinear,pekalski1998three,wilensky1997netlogo,wilensky1999netlogo}. We use $a_k$, $b_k$, and $c_k$ to denote the population sizes of species $A$, $B$, and $C$ at time $k$, respectively. In ecology, this model may describe the evolution of grass, sheep, and wolves, or plankton, forage fish, and predatory fish. In the case of Aspergillosis, the three species may represent iron (nutrient supply), Aspergillus (prey), and macrophages (predators)~\cite{oremland2016computational,ribeiro2022multi}. Generalized predator-prey models with even more species have found applications in studies of microbial communities~\cite{faust2012microbial}. 

We simulate the three-species predator-prey dynamics on a grid of size $L\times L$ with periodic boundaries. Each grid cell can be in one of two different states: (i) ``nutrient-rich'' and (ii) ``nutrient-poor''. Prey perform a random walk with a directional bias in positive $x$-direction and consume nutrients to stay alive. We use $\lambda_1$ to denote the energy gain per unit nutrient. Prey consume nutrients from the nearest available grid cell if it is in a ``nutrient-rich'' state. This grid cell is then switched to a ``nutrient-poor'' state and regenerates nutrients after $\tau$ periods. 

Predators also perform a random walk with a directional bias identical to that of prey, and they consume prey when both are located within the same grid cell. The energy gain per prey is $\lambda_2$. In each period, all predators and prey lose one unit of energy to sustain their metabolism. Predators and prey die if their energy levels fall below 0. They reproduce at rates $\alpha_{1}$ and $\alpha_{2}$, respectively. 

In Figure~\ref{fig:abm}, we show a snapshot of a three-species predator-prey ABM with $51\times51$ grid cells. Green and light brown grid cells represent nutrient-rich and nutrient-poor regions, respectively.

\begin{figure*}[t]
    \centering
    \includegraphics{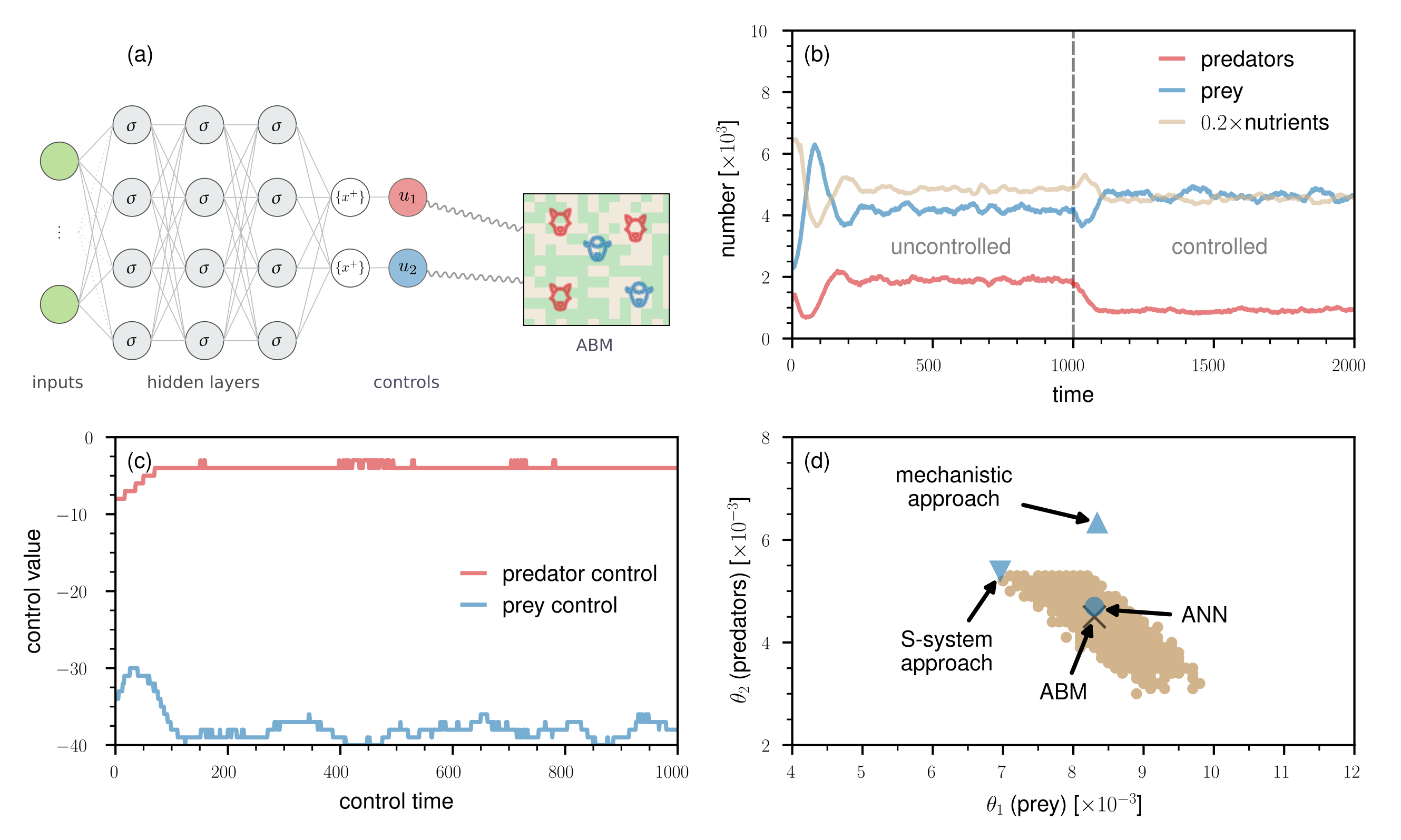}
    \caption{Control of predator-prey dynamics with an ANN. (a) To control a predator-prey ABM, we need to define suitable inputs and outputs for the ANN. Potential inputs are the population sizes $a_k$, $b_k$, and $c_k$ of species $A$, $B$, and $C$ at time $k$. We aim at directly managing the numbers of predator and prey, so there are two outputs $u_1$ and $u_2$. Using a problem-tailored straight-through estimator, the ANN outputs nonnegative integer-valued controls $u_1,u_2$ after subtracting the fractional part $\{[\cdot]^+\}$ of the positive part of the hidden-layer outputs. We use $\sigma$ and $\{x^+\}$ to indicate hidden-layer activations and the straight-through estimator, respectively. (b) The evolution of nutrient-rich lattice sites, prey, and predators. The vertical dashed grey line indicates the time at which the ANN controller is switched on.  The controller aims at increasing the mean number of prey by 10\% and reducing the mean number of predators by 50\%. We used a $255\times 255$ grid and set $b_0=2500$, $c_0=1250$, $\alpha_1=4.0$, $\alpha_2=5.0$, $\lambda_1=4.0$, $\lambda_2=20.0$, and $\tau=30$~\cite{wilensky1997netlogo}. The initial proportion of nutrient-rich lattice sites is 50\%. (c) The control outputs $u_1(b_k;\theta_1)$ (\ie, prey control) and $u_2(c_k;\theta_2)$ (\ie, predator control) as a function of the control time. (d) The values of $\theta_1$ and $\theta_2$ learned by different control methods (blue disk: ANN controller; blue triangle: mechanistic approach; blue inverted triangle: S-system approach). The black cross indicates the optimal values of the parameters $\theta_1,\theta_2$ found via a grid search and the orange dots indicate corresponding 1$\sigma$-intervals.}
    \label{fig:predator_prey}
\end{figure*}
To control the predator-prey dynamics, we need to define suitable inputs and outputs for the ANN. Potential inputs are the population sizes $a_k$, $b_k$, and $c_k$ of species $A$, $B$, and $C$ at time $k$. In the two control problems that we consider in this paper, we aim at directly managing the numbers of predator and prey, so there are two outputs $u_1$ and $u_2$ [see Figure~\ref{fig:predator_prey}(a)]. In the first control problem, our objective is to steer the dynamics towards a new stable state characterized by more prey and fewer predators. In the second control problem, we focus on controlling the dynamics during a transient phase. The ABM that we use in our simulations has $255\times 255$ grid cells.
\subsubsection*{Steady-state control}
Before controlling the considered ABM, we let it evolve for $1000$ time steps to estimate the steady-state numbers of predators and prey. For the parameters that we use in our simulations, the mean numbers of predators and prey over the final $100$ time steps are about $1896$ and $4159$, respectively. When controlling the dynamics, we allow it to evolve for an additional $1000$ time steps. We use $N_t$ to denote the total number of time steps, which is $2000$ in this example. 

Our control target is to increase the mean number of prey in the last $N^*_t$ time steps of the controlled time horizon by 10\% and reduce the corresponding mean number of predators by 50\% [see Figure~\ref{fig:predator_prey}(b)]. Intuitively, such a large reduction in the mean number of predators is associated with a large increase in the number of prey. Hence, the control function that we wish to identify has to reduce both the number of prey and predators in the steady state of the ABM. This can be achieved with a two-node ANN controller in which the two outputs are used to control the population sizes of both predators and prey (see Materials and Methods). To train the ANN controller, we use the quadratic loss function
\begin{equation}
    J_1(\boldsymbol{\theta})=\big(\bar{b}(\boldsymbol{\theta})-\bar{b}^*\big)^2+\big(\bar{c}(\boldsymbol{\theta})-\bar{c}^*\big)^2\,,
\label{eq:loss}
\end{equation}
where $\boldsymbol{\theta}\in\mathbb{R}^N$ denote ANN parameters, and $\bar{b}^*$ and $\bar{c}^*$ are the desired target states (\ie, the desired mean number of prey and predators over the last $N^*_t$ time steps). In this first, steady-state control example, we have $N=2$ ANN parameters (\ie,  $\boldsymbol{\theta}=(\theta_1,\theta_2)^\top$), $\bar{b}^*=4575$, $\bar{c}^*=948$, and $N^*_t=100$. The quantities
\begin{align}
    \bar{b}(\boldsymbol{\theta})=\frac{1}{N^*_t}\sum_{k=1}^{N^*_t} b_{(N_t-N^*_t+k)}(\boldsymbol{\theta})
\end{align}
and
\begin{align}
    \bar{c}(\boldsymbol{\theta})=\frac{1}{N^*_t}\sum_{k=1}^{N^*_t} c_{(N_t-N^*_t+k)}(\boldsymbol{\theta})
\end{align}
are the corresponding reached states. We parameterize the integer-valued control function $\mathbf{u}(b_k,c_k;\boldsymbol{\theta})$ according to
\begin{equation}
    \mathbf{u}(b_k,c_k;\boldsymbol{\theta}) = 
    \begin{pmatrix}
    -([b_k\theta_1]^+-\{[b_k\theta_1]^+\}) \\
    -([c_k\theta_2]^+-\{[c_k\theta_2]^+\})
    \end{pmatrix}\,,
    \label{eq:control_two_node}
\end{equation}
where $[x]^+=\mathrm{ReLU}(x)=\max(0,x)$. The notation $\{x\}$ denotes the fractional part of $x$. That is, $\{x\}=x - \lfloor x \rfloor$ if $x>0$ and $\lfloor \cdot \rfloor$ denotes the floor function.\footnote{Using the control function specified in \eqref{eq:control_two_node} involves employing a problem-tailored straight-through estimator~\cite{asikis2021multi,bottcher2023control}, which enables the training of ANN controllers with integer-valued outputs through backpropagation.}

We use the two control signals $u_1(b_k;\theta_1)=-([b_k\theta_1]^+-\{[b_k\theta_1]^+\})$ and $u_2(c_k;\theta_2)=([c_k\theta_2]^+-\{[c_k\theta_2]^+\})$ to manage the population sizes of prey and predators, respectively. The control function \eqref{eq:control_two_node} is set up such that it outputs negative integer-valued controls, meaning that a certain number of prey and predators will be removed from the ABM at each time step. More details on the training of this controller are provided in the Materials and Methods.

The smallest loss $J_1(\boldsymbol{\theta})$ of about 74.09 achieved during training is associated with the parameters $\theta_1=0.0083$ and $\theta_2=0.0047$. The corresponding numbers of reached prey and predators are $\bar{b}(\boldsymbol{\theta})\approx 4573$ and $\bar{c}(\boldsymbol{\theta})\approx 956$. 

In Figure~\ref{fig:predator_prey}(c), we show the evolution of $u_1(b_k;\theta_1)$ (\ie, prey control) and $u_2(c_k;\theta_2)$ (\ie, predator control) within the control horizon. At each time step, around 3--4 predators and between 35--40 predators are removed.

The learned ANN parameters $(\theta_1,\theta_2)=(0.0083,0.0047)$ (blue disk) are close to the optimal ones $(0.0083,0.0045)$ (black cross) [see Figure~\ref{fig:predator_prey}(d)]. We determined the optimal parameters for prey ($\theta_1$) and predators ($\theta_2)$ by performing a grid search over the underlying parameter space. All control solutions found to be one standard deviation away from the optimum are colored in brown. This brown region is a result of the stochasticity that is present in the considered ABM dynamics. For comparison, in Figure~\ref{fig:predator_prey}(d), we also show the values of $\theta_1,\theta_2$ as identified by two ODE metamodeling approaches that have been recently proposed in \cite{fonsecametamodels2024}. One metamodel is based on a mechanistic Lotka--Volterra approximation of the ABM while the second metamodel uses an S-system approach that is rooted in biochemical systems theory~\cite{savageau1969iochemical,savageau1970biochemical,voit2013biochemical}. In the Materials and Methods, we provide further details on the two metamodels. The solutions associated with both the mechanistic metamodel (blue triangle) and the S-system approach (blue inverted triangle) are more distant from the optimum compared to the solution based on the ANN control approach. This observation holds true for the other metamodels studied in \cite{fonsecametamodels2024}. 

There are two steps involved in controlling an ABM with a metamodel. First, the metamodel has to be trained to approximate ABM dynamics. Second, a solution to a given control problem that has been found using a metamodel will be lifted back to the ABM. Depending on their parameterization, metamodels may not fully capture the complex behavior of an ABM; as a result, control solutions based on metamodels may deviate from the desired ABM control. In contrast, the employed ANN controller directly operates on the ABM and can achieve better solutions than the ODE metamodels considered in \cite{fonsecametamodels2024}. While the ANN controller we used in this example can achieve better solutions than metamodel-based control, not all ABM control problems can be directly addressed with this method. As an example, we will consider a control problem associated with a metabolic network ABM~\cite{fonsecametamodels2024} in the last part of the Results. Instead of directly controlling the metabolic network with an ANN, we will show how to solve the control problem with a neural ODE~\cite{wang1998runge,DBLP:conf/nips/ChenRBD18} metamodelling approach.
\subsubsection*{Transient control}
\begin{figure}[t]
    \centering
\includegraphics{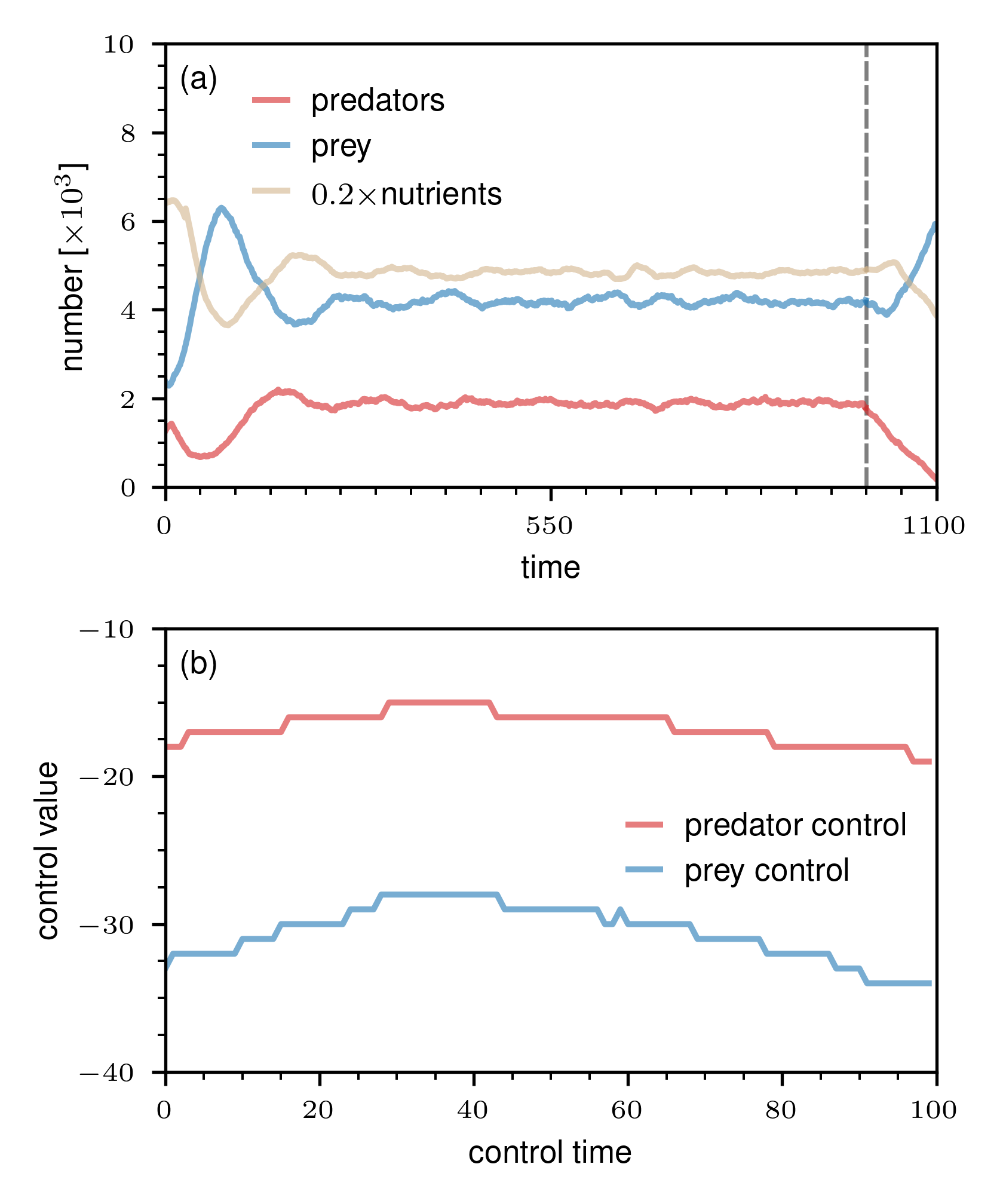}
    \caption{Control of transient predator-prey dynamics with an ANN. (a) The evolution of nutrient-rich lattice sites, prey, and predators. The vertical dashed grey line indicates the time at which the ANN controller is switched on. The controller aims at increasing the mean number of prey by 10\% and reducing the mean number of predators by 50\%. We used a $255\times 255$ grid and set $b_0=2500$, $c_0=1250$, $\alpha_1=4.0$, $\alpha_2=5.0$, $\lambda_1=4.0$, $\lambda_2=20.0$, and $\tau=30$. The initial proportion of nutrient-rich lattice sites is 50\%. (b) The control outputs $u_1(b_k;\boldsymbol{\theta})$ (\ie, prey control) and $u_2(c_k;\boldsymbol{\theta})$ (\ie, predator control) as a function of the control time.}
    \label{fig:predator_prey_transient}
\end{figure}
Before focusing on the metabolic network model, we briefly examine the application of the direct ANN control method that we employed in the prior section to transient dynamics. As in the steady-state control example, our goal is to increase the mean number of prey by 10\% and reduce the mean number of predators by 50\%. However, instead of considering a long control time horizon of $1000$ time steps during which the dynamics is steered into a new steady state, we now want to adjust the mean numbers of the two species over a shorter time period of $100$ time steps. To equip the ANN controller with a higher representational capacity, we use an ANN with three hidden layers and between 16--64 neurons per layer. More details on the ANN structure and training are provided in the Materials and Methods.

In Figure~\ref{fig:predator_prey_transient}(a), we show an example of controlled transient dynamics. The loss $J_1(\boldsymbol{\theta})$ associated with the underlying controller is about 28.00. The corresponding numbers of reached prey and predators are $\bar{b}(\boldsymbol{\theta})\approx 4576$ and $\bar{c}(\boldsymbol{\theta})\approx 953$, respectively. Examining Figure~\ref{fig:predator_prey_transient}(b), we observe that the magnitude of the predator control signal $u_2(c_k;\boldsymbol{\theta})$ is substantially larger than in the earlier steady-state control example [see Figure~\ref{fig:predator_prey}(c)]. On the other hand, the magnitude of the prey control signal $u_1(c_k;\boldsymbol{\theta})$ is smaller than in the prior example. 

We also examined the performance of the transient-dynamics ANN controller on 50 unseen test samples. The corresponding mean numbers of prey and predators were found to be 4485($\pm 98$) and 966($\pm 46$), respectively.\footnote{Values in parentheses indicate the unbiased sample standard deviation.} These results indicate that the controller performs well on unseen samples.
\subsection*{Metabolic pathway model}
\begin{figure}[t]
    \centering
\includegraphics{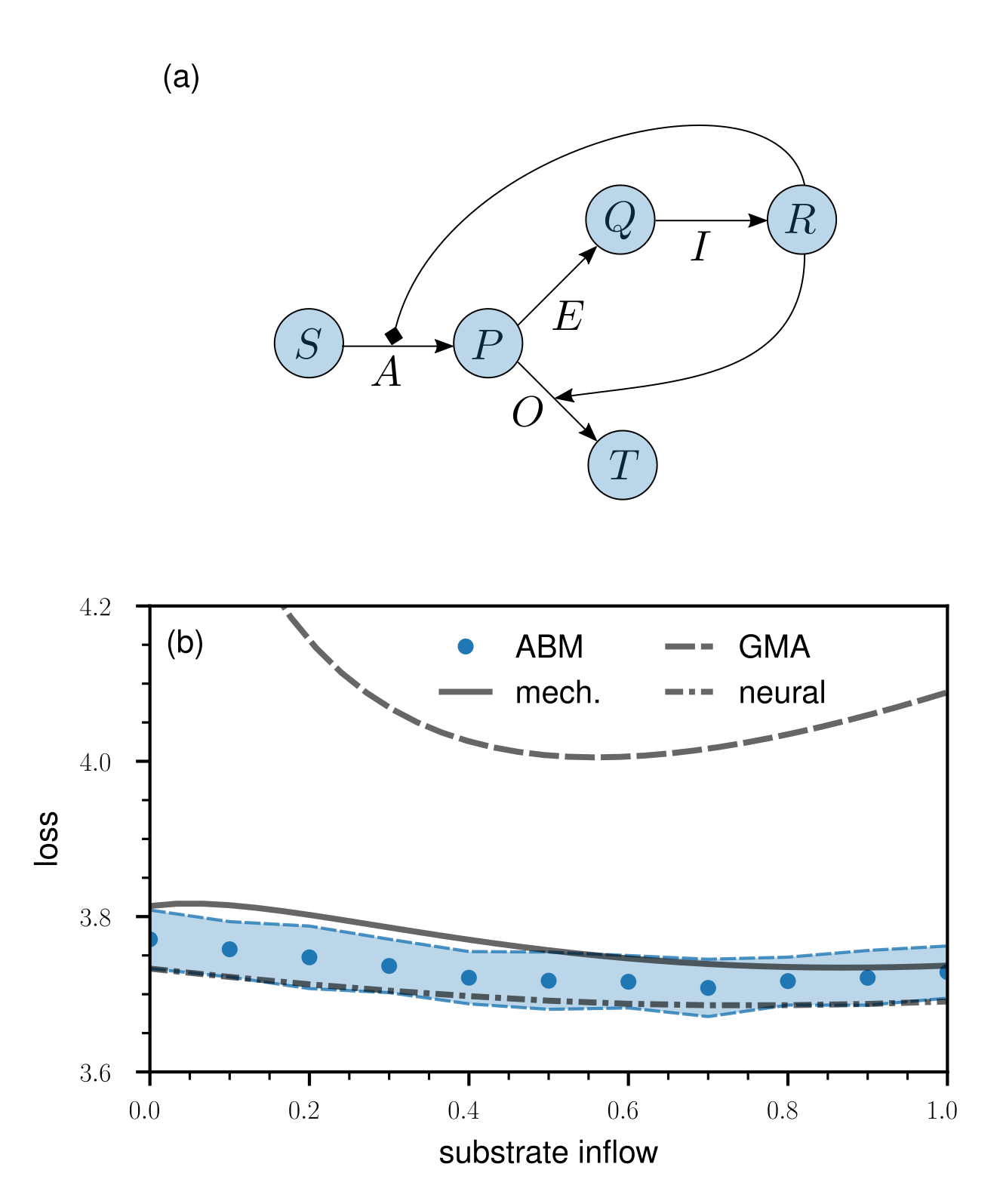}
    \caption{Learning and controlling metabolic pathway dynamics with neural ODEs. (a) Overview of the reactions in the metabolic pathway model. There are four reactions associated with five metabolites ($S$, $P$, $Q$, $R$, and $T$) and four enzymes ($A$, $E$, $I$, and $O$). The two arrows originating from metabolite $R$ indicate that it inhibits enzyme $A$ and increases the rate of enzyme $O$. In our ABM simulations, all reactions are modelled at the microscale level. The initial amounts of metabolites $S$, $P$, $Q$, $R$, and $T$ are $8\times 10^4$, $2\times 10^4$, $2\times 10^4$, $10$, and $10$, respectively. The initial amount of each of the four enzymes is $200$. (b). The loss of the metabolic pathway control problem [see \eqref{eq:loss_metabolic}] as a function of the inflow of substrate $S$ [blue dots: ABM; solid line: mechanistic metamodel; dashed line: generalized mass action (GMA) metamodel; dash-dotted line: neural ODE metamodel]. The blue-shaded regions indicate $1\sigma$-intervals that are based on 100 ABM instantiations. Minimizing the loss function means minimizing substrate depletion and maximizing the production of reaction products.}
    \label{fig:metabolic_network}
\end{figure}

We will now focus on a second ABM control problem for which the direct control approach used in the previous example cannot be implemented straightforwardly. The control problem under consideration involves a metabolic network that is based on four reactions associated with five metabolites [see Figure~\ref{fig:metabolic_network}(a)]. We model all interactions between enzymes, metabolites, and their respective complexes at the microscale level.

In this ABM, we employ three types of agents: (i) metabolites, (ii) enzymes, and (iii) enzymatic complexes. In total, there are five metabolites, four enzymes, and 12 enzyme-metabolite complexes.\footnote{The ABM incorporates more than 30 parameters, and their values are provided in the code repository accompanying this submission.} Metabolites in our model move ten times faster than enzymes and complexes. When metabolites are in proximity to enzymes or complexes, they may bind. Additionally, complexes may dissociate into their components at any time.

Enzymes have the capability to create complexes with their corresponding substrates, products, and regulators. Furthermore, enzyme-substrate complexes can undergo catalysis, resulting in the formation of a complex between the enzyme and the product. We treat all four enzymatic reactions as irreversible.

The control problem that we wish to solve aims at identifying the optimal substrate inflow to minimize substrate waist and maximize the generation of reaction products. Mathematically, our objective is to determine the constant inflow of substrate, $q\in[0,1]$ per time step, that minimizes the loss function
\begin{equation}
    J_2(q)=\frac{\sum_{k=1}^{N_t} S_k}{\sum_{k=1}^{N_t} R_k+T_k}\,,
    \label{eq:loss_metabolic}
\end{equation}
where $S_k$ denotes the concentration of substrate at time step $k$, while $R_k$ and $T_k$ denote the concentrations of the corresponding end-products of the pathway. In all simulations, we set $N_t=5\times 10^4$. The initial amounts of metabolites $S$, $P$, $Q$, $R$, and $T$ are $8\times 10^4$, $2\times 10^4$, $2\times 10^4$, $10$, and $10$, respectively. The initial amount of each of the four enzymes is $200$.
\subsubsection*{Neural ODE metamodel and controller}
A direct application of neural-network controllers as in the predator-prey ABM is challenging because of the various reactions that one would have to consider when keeping track of the effect of control inputs on the metabolic dynamics during training. An alternative is provided by the metamodeling approach that has been proposed in \cite{an2017optimization,fonsecametamodels2024}. The basic idea is to first identify control signals in ODE metamodels and then lift them back to an ABM. For the metabolic network model, both a mechanistic Michaelis--Menten approximation and generalized mass action (GMA) models~\cite{savageau1969iochemical,savageau1970biochemical,voit2013biochemical} provide good descriptions of the underlying reactions. We define both metamodels and describe their training in the Materials and Methods.

Based on the data that we show in Figure~\ref{fig:metabolic_network}(b), we conclude that both metamodels are valuable for obtaining estimates of the optimal substrate inflow. The mechanistic metamodel and the ABM exhibit closely aligned loss values, whereas the GMA metamodel shows substantially larger losses than those of the ABM. Despite this, the GMA approximation provides a slightly more accurate estimate of approximately 0.6 for the optimal ABM substrate inflow, which is around 0.7. In contrast, the mechanistic metamodel suggests an optimal substrate inflow of about 0.9, a value slightly farther from the ABM optimum compared to the GMA estimate.

A complementary approach that does not require one to manually set up ODE metamodels is based on neural ODEs, which have found applications in system identification and control~\cite{wang1998runge,DBLP:conf/nips/ChenRBD18,lagergren2020biologically,fronk2023interpretable,bottcher2022ai,asikis2022neural,bottcher2022near,mowlavi2023optimal,bottcher2023control,bottcher2023gradient}. We train a neural ODE metamodel on ABM instances that are based on the same values of substrate inflow as in the two other metamodels (see Materials and Methods for further details). We then used the trained neural ODE model to determine the optimal substrate inflow that minimizes the loss $J_2(q)$ [see \eqref{eq:loss_metabolic}]. The neural ODE identifies an optimal substrate inflow of 0.7, which coincides with the optimum of the ABM.\footnote{The number of parameters in the neural ODE metamodel is 120. In the mechanistic and GMA metamodels, the number of parameters are 17 and and 24, respectively.} Neural ODE metamodels can thus provide a valuable alternative to other metamodels when no or only very little mechanistic information is available about the underlying ABM.
\subsubsection*{Combining a mechanistic metamodel with a neural ODE controller}
Neural ODEs can be also used to directly parameterize control functions. As an example, we combine the mechanistic metamodel of the metabolic pathway dynamics with a neural ODE controller that outputs a time-dependent substrate inflow. The corresponding loss in the metamodel is 3.61. This value is smaller than the minimum loss of 3.73 obtained with the optimal constant substrate inflow [see Figure~\ref{fig:metabolic_network}(b)]. To test if these loss improvements are achievable in the metabolic pathway ABM, we fed the output of the time-dependent neural ODE controller into the ABM. The corresponding ABM loss is 3.60($\pm 0.03$), which is smaller than the minimum loss of 3.71$(\pm0.04)$ achieved with a constant substrate inflow [see Figure~\ref{fig:metabolic_network}(b)]. 

\begin{figure}
    \centering
    \includegraphics{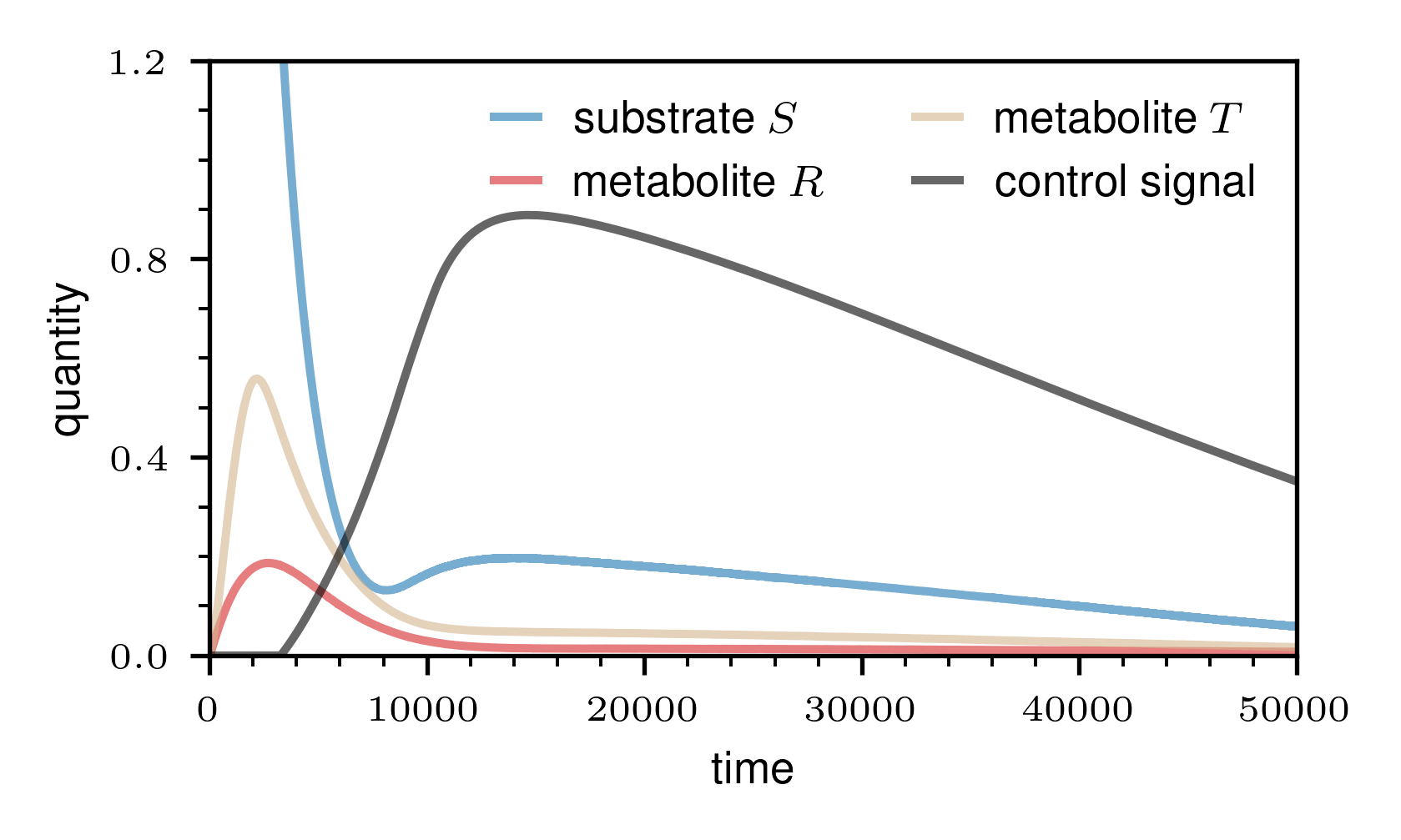}
    \caption{Controlling metabolic pathway dynamics with a neural ODE controller. We show the evolution of the amount of substrate $S$ , the amounts of metabolites $R$ and $T$, and the control signal. The control signal is generated by a neural ODE controller, which we trained using a mechanistic Michaelis--Menten metamodel and the loss as defined in \eqref{eq:loss_metabolic}. The shown amounts of $S$, $R$, and $T$ are averages over 100 ABM instantiations. The initial amounts of metabolites $S$, $R$, and $T$ are $8\times 10^4$, $10$, and $10$, respectively. In the shown plot, we rescaled these quantities by $10^{-4}$.}
    \label{fig:metabolic_time_dependent}
\end{figure}

In Figure~\ref{fig:metabolic_time_dependent}, we show the evolution of the amount of substrate $S$, the amounts of metabolites $R$ and $T$, and the time-dependent neural ODE controller output $q_k(\boldsymbol{\theta})$. The initial amount of $S$ is high compared to those of $R$ and $T$. Minimizing the loss defined in \eqref{eq:loss_metabolic} means that we have to minimize the total amount of substrate $\sum_k S_k$ divided by the total amount of reaction products $\sum_k R_k + T_k$. Instead of adding substrate to the system in the beginning, as done with a constant inflow rate $q$, the neural ODE controller learned that adding substrate in the first few thousand time steps is not needed to achieve good loss values [see Figure~\ref{fig:metabolic_time_dependent}]. The inflow of substrate $q_k(\boldsymbol{\theta})$ gets larger as the concentration of $S$ approaches values close to those of $R$ and $T$. 

In summary, neural ODEs are valuable not only as metamodels but also as control functions that can be seamlessly integrated with the former.
\section*{Discussion}
Control problems commonly arise in various biomedical contexts such as treatment design and pharmacology. However, conducting direct comparisons and optimizations of different treatments \emph{in vivo} is often impractical. Rather than relying primarily on laboratory and clinical trials, medical digital twins offer a complementary approach for testing and developing treatments through \emph{in silico} optimization. 

While a diverse array of models can be integrated into medical digital twins, ABMs are among the most commonly used model types for simulating heterogeneous, multi-species biomedical systems. Despite their increasing use in biomedicine, there currently exists no general methodology to control ABMs. For example, traditional control theory methods, which have been developed for ODE systems, are not directly applicable to ABMs.

Metamodels provide a useful approach in connecting traditional control theory with ABMs~\cite{an2017optimization, fonsecametamodels2024}. The main idea behind metamodeling involves training mechanistically inspired or more general ODE systems on ABM dynamics and solving control problems within these ODE systems. Once an appropriate control solution is identified, it is then applied back to the ABM. Similar to other function approximators, metamodels are also subject to the bias-variance tradeoff. High-bias mechanistic approximations may be beneficial when detailed information about the inner workings of a given ABM is available, whereas lower-bias models, such as S-system approaches~\cite{savageau1969iochemical,savageau1970biochemical,voit2013biochemical}, may be more suitable for ABMs for which such details are unknown.

In this paper, we developed complementary ABM-control methods that use artificial neural networks (ANNs) as control functions. We first considered a predator-prey ABM, where we managed the number of predators and prey by directly controlling the ABM with an ANN controller. We addressed two control tasks: (i) steady-state control and (ii) transient control. In both tasks, the ANN controller successfully identified suitable control signals. For the steady-state control task, a comparison between the metamodel-based controls proposed in \cite{fonsecametamodels2024} and the ANN controls revealed that the ANN can identify control solutions much closer to the optimum because it directly operates on the ABM without using any approximations.

In a second ABM describing metabolic pathway dynamics, we addressed a control problem aimed at determining the optimal substrate inflow to minimize substrate depletion and maximize the generation of reaction products. For this system, we employed neural ODE metamodels which we then used to identify suitable control signals. We found that the neural ODE approach was able to compete favorably with the best metamodels that have been proposed in \cite{fonsecametamodels2024}. We also showed that neural ODEs can be integrated into existing metamodels to learn effective control functions.

Our findings suggest that ANN controllers are valuable in addressing different ABM control problems. The ability of ANNs to act as universal function approximators~\cite{hornik1991approximation,hanin2017approximating,park2020minimum}, combined with advancements in automatic differentiation and optimizer development, renders them well-suited for solving biomedical control problems. However, it is important to emphasize that our work should be seen as just an initial step towards solving intricate optimization and control problems associated with medical digital twins. More research is necessary to connect the proposed and related control approaches to medical digital twins, especially concerning models that are dynamically updated with patient data. 

Another worthwhile direction for future work is to provide uncertainty quantification and further insights into the generalization behavior of neural ODE metamodels \cite{hammouri2023uncertainty,DBLP:conf/cdc/CheeHP23}. For instance, in a mechanistic metamodel, we can anticipate effective generalization across a relatively broad parameter range. However, neural-ODE and non-mechanistic metamodels may face limitations in capturing the behavior of a medical digital twin under parameter changes. Additionally, unlike mechanistic metamodels, those based on neural ODEs may exhibit a higher susceptibility to overfitting. Therefore, integrating mechanistic information into a neural ODE is useful to introduce a suitable inductive bias into the learning process.

In summary, research at the intersection of machine learning, biomedicine, and control theory can be expected to not only contribute to enhance our ability to develop more effective treatments but also holds the potential to help address challenging control problems in related fields.
\subsection*{Data Archival}
Our source codes are publicly available at \url{https://gitlab.com/ComputationalScience/abm-control}.
\clearpage
\onecolumngrid
\section*{Materials and Methods}
\subsection*{Neural-network architectures and training}
\subsubsection*{Steady-state controller (predator-prey dynamics)}
The ANN architecture of the steady-state controller has two inputs: (i) the number of species of type $B$ at time $k$, $b_k$, and (ii) the number of species of type $C$ at time $k$, $c_k$. It has two outputs $u_1(b_k;\theta_1)$ and $u_2(c_k;\theta_2)$, which are used to control the population sizes of species $B$ and $C$, respectively. The controller consists of two neurons with ReLU activation function and two weights $\theta_1,\theta_2$.

We train the ANN for 50 epochs using RMSProp and a learning rate of $10^{-4}$. Each training epoch takes about 3.5 minutes one Ryzen threadripper 3970x CPU core with a 3.7 GHz clock speed. The initial weights are both set to $5\times 10^{-3}$.

To train the ANN using backpropagation and still output integer-valued quantities, we employ a problem-tailored straight-through estimator~\cite{asikis2021multi,bottcher2023control}. In a straight-through estimator, a mathematical operation that is applied in a forward pass is treated as an identity operation in the backward pass. In accordance with \cite{asikis2021multi,bottcher2023control}, we implement a straight-through estimator that rounds neural-network outputs by subtracting from the positive parts $[y]^+$ of the outputs $y$ of the last hidden layer the corresponding fractional parts $\{[y]^+\}$. That is, in a forward pass, the neural network output is $[y]^+ - \{[y]^+\}$, where $\{y\}=y - \lfloor y \rfloor$ if $y>0$ and $\lfloor \cdot \rfloor$ denotes the floor function. While updating neural-network weights by backpropagating gradients, we detach the fractional part $\{[y]^+\}$ from the underlying computational graph such that the neural-network output is treated as $[y]^+$. 
\subsubsection*{Transient controller (predator-prey dynamics)}
The ANN architecture that we use to control transient predator-prey dynamics consists of three fully connected hidden layers with 64, 32, and 16 continuously differentiable exponential linear unit (CELU) activations each. Using CELU functions offers an advantage over traditional rectified linear units (ReLUs) because the former are continuously differentiable and produce non-zero outputs for negative arguments. This feature helps prevent the occurrence of the ``dead ReLU'' problem, associated with near-zero outputs and vanishingly small gradients. Mathematically, the CELU activation is
\begin{equation}
{\rm CELU}(x,\alpha)=\left[x\right]^+ - \left[\alpha \left(1-\exp(x/\alpha)\right)\right]^+\,,
\label{eq:celu}
\end{equation}
which approaches ${\rm ReLU}(x)=[x]^+$ in the limit $\alpha\rightarrow 0^+$~\citep{barron2017continuously}. In this work, we set $\alpha=1$.

The transient-dynamics ANN controller has two inputs: (i) the number of species of type $B$ at time $k$, $b_k$, and (ii) the number of species of type $C$ at time $k$, $c_k$. It has two outputs ${u}_1(b_k,c_k;\boldsymbol{\theta})$ and ${u}_2(b_k,c_k;\boldsymbol{\theta})$, which are used to control the population sizes of species $B$ and $C$, respectively.

We train the ANN using RMSProp and a learning rate of $10^{-4}$. The initial weights and biases are set to $10^{-2}$. As in the steady-state controller, we use the same straight-through estimator. Loss values of about 10--100 can be achieved after 50--100 training epochs. One training epoch takes about 20 seconds on one Ryzen threadripper 3970x CPU core with a 3.7 GHz clock speed.

To avoid that the controller produces too large control signals in the first few training epochs, we pre-train the ANN, minimizing
\begin{align}
\begin{split}
\tilde{J}_1(\boldsymbol{\theta})&=\left[\left(\textstyle\sum_{k=1}^{10} {u}_1(b_k,c_k;\boldsymbol{\theta})\right)^2-c_1\right]^2\\
&+\left[\left(\textstyle \sum_{k=1}^{10} {u}_2(b_k,c_k;\boldsymbol{\theta})\right)^2-c_2\right]^2\,,
\end{split}
\end{align}
where we set $c_1=10^4$ and $c_2=10^5$. As controller inputs in the pre-training phase, we use the steady-state values of the uncontrolled dynamics (\ie, $b_k=4159$ and $c_k=1896$ for $k\in\{1,\dots,10\}$). 
\subsection*{ODE metamodels}
As baselines for the ANN controllers that we study in this work, we use controls identified by two ODE metamodels that have been proposed in \cite{fonsecametamodels2024}. The first metamodelling approach uses a mechanistic approximation while the second one is based on an S-system approximation that is rooted in biochemical systems theory~\cite{savageau1969iochemical,savageau1970biochemical,voit2013biochemical}. 

In the following sections, we will provide a brief overview of the mathematical forms of the metamodels considered in this study. For further details on the training of ODE metamodels that are not based on neural ODEs, we refer the reader to \cite{fonsecametamodels2024}.
\subsubsection*{Mechanistic approach (predator-prey dynamics)}
For the predator-prey ABM, the mechanistic metamodel is given by
\begin{equation}
    \begin{pmatrix}
    \dot{X} \\
    \dot{Y} \\
    \dot{Z}
    \end{pmatrix} = 
    \begin{pmatrix}
     p_1 X-p_2 X^2-p_3 X Y \\
      p_4 X Y-p_5 Y-p_6 Y Z  \\
      p_7 Y Z-p_8 Z
    \end{pmatrix}-
    \begin{pmatrix}
    0\\
    \theta_1 Y\\
    \theta_2 Z
    \end{pmatrix}\,,
    \label{eq:mec_ode}
\end{equation}
where $X\equiv X(t)$, $Y\equiv Y(t)$, and $Z\equiv Z(t)$ denote the numbers of prey, predators, and nutrients at time $t$, respectively. It is a three-species Lotka--Volterra model with an additive control term. The model parameters are $p_i$ ($i\in\{1,\dots,8\}$) and $\theta_1,\theta_2$ are the control parameters that we employ to manage the number of prey and predators. We included the model parameters in the code repository accompanying this submission.
\subsubsection*{S-system approach (predator-prey dynamics)}
While the three-species Lotka--Volterra model in \eqref{eq:mec_ode} has eight parameters (without counting the two control parameters $\theta_1,\theta_2$), the S-system metamodel involves 24 parameters $p_i$ ($i\in\{1,\dots,24\}$). It is given by
\begin{equation}
    \begin{pmatrix}
    \dot{X} \\
    \dot{Y} \\
    \dot{Z}
    \end{pmatrix} = 
    \begin{pmatrix}
     p_1 X^{p_7} Y^{p_{13}} Z^{p_{19}}-p_2 X^{p_8} Y^{p_{14}} Z^{p_{20}} \\
     p_3 X^{p_{9}} Y^{p_{15}} Z^{p_{21}}-p_{4} X^{p_{10}} Y^{p_{16}} Z^{p_{22}} \\
     p_{5} X^{p_{11}} Y^{p_{17}} Z^{p_{23}}-p_{6} X^{p_{12}} Y^{p_{18}} Z^{p_{24}}
    \end{pmatrix}-
    \begin{pmatrix}
    0\\
    \theta_1 Y\\
    \theta_2 Z
    \end{pmatrix}\,.
    \label{eq:SS_ode}
\end{equation}
The control term is the same as in the mechanistic approach. We included the model parameters in the code repository accompanying this submission.
\subsubsection*{Mechanistic approach (metabolic pathway dynamics)}
In the mechanistic metamodel of the metabolic pathway dynamics, we approximate the dynamics of the five metabolites $S\equiv S(t)$, $P\equiv P(t)$, $Q\equiv Q(t)$, $R\equiv R(t)$, and $T\equiv T(t)$ using a Michaelis-–Menten type rate law. The metamodel is
\begin{equation}
    \begin{pmatrix}
    \dot{S} \\
    \dot{P} \\
    \dot{Q} \\
    \dot{R} \\
    \dot{T}
    \end{pmatrix} = 
    \begin{pmatrix}
    q \\
    0 \\
    0 \\
    0 \\
    0
    \end{pmatrix} + 
    \begin{pmatrix}
     -1 & 0 & 0 & 0 \\
     1 & -1 & 0 & -1 \\
     0 & 1 & -1 & 0 \\
     0 & 0 & 1 & 0 \\
     0 & 0 & 0 & 1
    \end{pmatrix}
    \begin{pmatrix}
     F_A \\
     F_E \\
     F_I \\
     F_O
    \end{pmatrix}-k
    \begin{pmatrix}
    S \\
    P \\
    Q \\
    R \\
    T
    \end{pmatrix}\,,
    \label{eq:Mec_EnzKin}
\end{equation}
where
\begin{equation}
\begin{split}
       F_A &= \frac{p_1\frac{S}{p_2}}{1+\frac{S}{p_2}+\frac{P}{p_3}+\frac{R}{p_4}}\,,\\
    F_E &= \frac{p_5\frac{P}{p_6}}{1+\frac{P}{p_6}+\frac{Q}{p_7}} \,,\\
    F_I &= \frac{p_8\frac{Q}{p_9}}{1+\frac{Q}{p_9}+\frac{R}{p_{10}}} \,,\\
    F_O &= \frac{p_{11}\frac{P}{p_{12}}+p_{13}\frac{R}{p_{14}}\frac{P}{p_{15}}}{1+\frac{P}{p_{12}}+\frac{T}{p_{16}}+\frac{R}{p_{14}}\left(1+\frac{P}{p_{15}}+\frac{T}{p_{17}}\right)} \,,
\end{split}
    \label{eq:Mec_EnzKin_F}
\end{equation}
denote the fluxes through the reactions catalyzed by the enzymes $A$, $E$, $I$, and $O$. The quantity $q$ is the rate of inflow of substance $Q$ and $k$ is the metabolite-removal rate. The Michaelis--Menten approximation is associated with 17 parameters $p_i$ ($i\in\{1,\dots,17\}$). The parameters are included in the code repository accompanying this submission.
%

\subsubsection*{Generalized mass action approach (metabolic pathway dynamics)}
The generalized mass action (GMA) approach~\cite{savageau1969iochemical,savageau1970biochemical,voit2013biochemical} has the same mathematical structure as the mechanistic metamodel [see \eqref{eq:Mec_EnzKin}], but it allows for a more straightforward determination of fluxes. Instead of the mechanistically determined fluxes [see \eqref{eq:Mec_EnzKin_F}] with 17 parameters, we use the fluxes
\begin{equation}
    \begin{pmatrix}
    F_A \\
    F_E \\
    F_I \\
    F_O 
    \end{pmatrix} = 
    \begin{pmatrix}
    p_1 S^{p_{5}} P^{p_{9}} Q^{p_{13}} R^{p_{17}} T^{p_{21}} \\
    p_2 S^{p_{6}} P^{p_{19}} Q^{p_{14}} R^{p_{18}} T^{p_{22}} \\
    p_3 S^{p_{7}} P^{p_{11}} Q^{p_{15}} R^{p_{19}} T^{p_{23}} \\
    p_4 S^{p_{8}} P^{p_{12}} Q^{p_{16}} R^{p_{20}} T^{p_{24}}
    \end{pmatrix}
    \label{eq:GMA_F}
\end{equation}
with 24 parameters $p_i$ ($i\in\{1,\dots,24\}$) in the GMA metamodel. The parameters are included in the code repository accompanying this submission.
%
\subsubsection*{Neural ODE metamodel (metabolic pathway dynamics)}
In the neural ODE metamodel, we integrate mechanistic information concerning the substrate influx at a rate of $q$ and the outflow of all five metabolites at a rate of $k$. We model the remaining interactions between metabolites, enzymes, and complexes through a vector field $\mathbf{f}(\mathbf{x};\theta)$ generated by an ANN with parameters $\boldsymbol{\theta}\in\mathbb{R}^N$ and input $\mathbf{x}=(S,P,Q,R,T)^\top$. In this metamodel, the evolution of the metabolites is described by
\begin{equation}
    \begin{pmatrix}
    \dot{S} \\
    \dot{P} \\
    \dot{Q} \\
    \dot{R} \\
    \dot{T}
    \end{pmatrix} = 
    \begin{pmatrix}
    q \\
    0 \\
    0 \\
    0 \\
    0
    \end{pmatrix} + 
    \mathbf{f}(\mathbf{x};\boldsymbol{\theta})
    -k
    \begin{pmatrix}
    S \\
    P \\
    Q \\
    R \\
    T
    \end{pmatrix}\,.
    \label{eq:neural_ode}
\end{equation}
The ANN associated with $\mathbf{f}(\mathbf{x};\boldsymbol{\theta})$ consists of two fully connected hidden layers with five CELU activations each. Further details on the training are reported in the code repository accompanying this submission.
\subsection*{Combining a mechanistic metamodel with a neural ODE controller}
Instead of using a constant substrate inflow $q$ in the mechanistic metabolic pathway metamodel [see \eqref{eq:Mec_EnzKin}], another option is to consider a neural ODE controller that takes the current time step $k$ as an input and outputs a time-dependent substrate inflow $q_k(\boldsymbol{\theta})$, where $\boldsymbol{\theta}$ are the parameters of the ANN. The resulting evolution equation is
\begin{equation}
    \begin{pmatrix}
    \dot{S} \\
    \dot{P} \\
    \dot{Q} \\
    \dot{R} \\
    \dot{T}
    \end{pmatrix} = 
    \begin{pmatrix}
    q_k(\boldsymbol{\theta}) \\
    0 \\
    0 \\
    0 \\
    0
    \end{pmatrix} + 
    \begin{pmatrix}
     -1 & 0 & 0 & 0 \\
     1 & -1 & 0 & -1 \\
     0 & 1 & -1 & 0 \\
     0 & 0 & 1 & 0 \\
     0 & 0 & 0 & 1
    \end{pmatrix}
    \begin{pmatrix}
     F_A \\
     F_E \\
     F_I \\
     F_O
    \end{pmatrix}-k
    \begin{pmatrix}
    S \\
    P \\
    Q \\
    R \\
    T
    \end{pmatrix}\,.
    \label{eq:Mec_EnzKin_neural}
\end{equation}
The ANN associated with $q_k(\boldsymbol{\theta})$ consists of five fully connected hidden layers with five exponential linear units (ELU) each. We train the neural ODE controller using the loss function defined in \eqref{eq:loss_metabolic}. Subsequently, we use the trained controller as input in the metabolic pathway ABM. Further details on the ODE controller are reported in the code repository accompanying this submission.

\acknowledgements{LB acknowledges financial support from hessian.AI and the Army Research Office (grant number W911NF-23-1-0129). LLF and RL acknowledge financial support from the Defense Advanced Research Projects Agency (grant HR00112220038), and the National Institutes of Health (grants R01 GM127909 and R01 AI135128). RL also acknowledges financial support from the National Institutes of Health (grant R01 HL169974).}

\bibliographystyle{ieeetr}
\bibliography{refs}
\end{document}